\documentclass{aa}  
\usepackage{graphicx}
\usepackage{txfonts}
\usepackage{hyperref}
\begin{document} 

\title{Short Timescale Photometric and Polarimetric Behavior of two BL Lacertae Type Objects\thanks{Partly based on data obtained at the INAF / Telescopio Nazionale Galileo at the Canary Island of La Palma under program Id: A29TAC\_21 (PI: S. Covino).}}

\author{S. Covino\inst{\ref{in:Brera}},
	M.C. Baglio\inst{\ref{in:Insubria},\ref{in:Brera}},
	L. Foschini\inst{\ref{in:Brera}},
	A. Sandrinelli\inst{\ref{in:Insubria},\ref{in:Brera}},
	F. Tavecchio\inst{\ref{in:Brera}},
   	A. Treves\inst{\ref{in:Insubria},\ref{in:Bicocca}},
	H. Zhang\inst{\ref{in:Ohio},\ref{in:Los}},
	U. Barres de Almeida\inst{\ref{in:Rio}},
	G. Bonnoli\inst{\ref{in:Brera}},
	 M. B\"ottcher\inst{\ref{in:NW},\ref{in:Ohio}},
	M. Cecconi\inst{\ref{in:TNG}},
	F. D'Ammando\inst{\ref{in:BO},\ref{in:BOU}},
	L. di Fabrizio\inst{\ref{in:TNG}},
	M. Giarrusso\inst{\ref{in:CT}},
	F. Leone\inst{\ref{in:CT}},
	E. Lindfors\inst{\ref{in:Fin}},
	V. Lorenzi\inst{\ref{in:TNG}},
         E. Molinari\inst{\ref{in:TNG},\ref{in:Bassini}},
         S. Paiano\inst{\ref{in:PD}},
         E. Prandini\inst{\ref{in:CH}},         
         C.M. Raiteri\inst{\ref{in:Torino}},
         A. Stamerra \inst{\ref{in:Torino}}
         and
         G. Tagliaferri\inst{\ref{in:Brera}}
         }

\institute{INAF / Osservatorio Astronomico di Brera, Via Bianchi 46, 23807, Merate (LC), Italy\\
	\email{stefano.covino@brera.inaf.it}
	\label{in:Brera}
	\and
	Universit\`a degli Studi dell'Insubria - Via Valleggio 11, 22100 Como, Italy
	\label{in:Insubria}
	\and
         INFN Milano-Bicocca - Universit\`a degli Studi di Milano-Bicocca, Piazza della Scienza 3, 20126 Milano, Italy
         \label{in:Bicocca}
         \and
         Centro Brasileiro de Pesquisas F\'{\i}sicas, Rua Dr. Xavier Sigaud 150, Urca, Rio de Janeiro, RJ 22290-180, Brazil 
         \label{in:Rio}
         \and
       Astrophysical Institute, Department of Physics and Astronomy, Ohio University, Athens, OH\,45701, USA
      \label{in:Ohio}
      \and
      Theoretical Division, Los Alamos National Laboratory, Los Alamos, NM\,87545, USA
      \label{in:Los}
      \and
      Centre for Space Research, North-West University, Potchefstroom 2531, South Africa
      \label{in:NW}
      \and
         INAF / Fund. Galileo Galilei, Rambla Jos\'e Ana Fern\'andez Perez 7, 38712 Bre\~{n}a Baja (La Palma), Canary Islands, Spain
         \label{in:TNG}
      \and
      INAF / Istituto di Radioastronomia, 40129 Bologna, Italy
      \label{in:BO}
      \and
      DIFA, Università di Bologna, Viale Berti Pichat 6/2, 40127, Bologna, Italy
      \label{in:BOU}
      \and
      Dipartimento di Fisica e Astronomia, Universit\`a di Catania, Sezione Astrofisica, Via S. Sofia 78, 9512 Catania, Italy
      \label{in:CT}
      \and
      University of Turku and Department of Physics, University of Oulu, Finland
      \label{in:Fin}
      \and
      INAF / Istituto di Astrofisica Spaziale e Fisica Cosmica Milano, Via E. Bassini 15, 20133 Milano, Italy
      \label{in:Bassini}
      \and
      Universit\`a di Padova and INFN, 35131, Padova, Italy
      \label{in:PD}
      \and
      ISDC - University of Geneva, 1290, Versoix, Switzerland
      \label{in:CH}
      \and
        INAF / Osservatorio Astrofisico di Torino, Via Osservatorio 10, 10025, Pino Torinese, Italy\\
         \label{in:Torino}
      }

\date{}

\abstract
{Blazars are astrophysical sources whose emission is dominated by non-thermal processes, typically interpreted as synchrotron and inverse Compton emission. Although the general picture is rather robust and consistent with observations, many aspects are still unexplored.}
{Polarimetric monitoring can offer a wealth of information about the physical processes in blazars. Models with largely different physical ingredients can often provide almost indistinguishable predictions for the total flux, but usually are characterized by markedly different polarization properties. We explore, with a pilot study, the possibility to derive structural information about the emitting regions of blazars by means of a joint analysis of rapid variability of the total and polarized flux at optical wavelengths.}
{Short timescale (from tens of seconds to a couple of minutes) optical linear polarimetry and photometry for two blazars, \object{BL\,Lacertae} and \object{PKS\,1424+240}, was carried out with the PAOLO polarimeter at the 3.6\,m Telescopio Nazionale Galileo. Several hours of almost continuous observations were obtained for both sources.}
{Our intense monitoring allowed us to draw strongly different scenarios for \object{BL\,Lacertae} and \object{PKS\,1424+240}, with the former characterized by intense variability on time-scales from hours to a few minutes and the latter practically constant in total flux. Essentially the same behavior is observed for the polarized flux and the position angle. The variability time-scales turned out to be as short as a few minutes, although involving only a few percent variation of the flux. The polarization variability time-scale is generally consistent with the total flux variability. Total and polarized flux appear to be essentially uncorrelated. However, even during our relatively short monitoring, different regimes can be singled out. }
{No simple scenario is able to satisfactorily model the very rich phenomenology exhibited in our data. Detailed numerical simulations show that the emitting region should be characterized by some symmetry, and the inclusion of turbulence for the magnetic field may constitute the missing ingredient for a more complete interpretation of the data.}
{}

\keywords{Polarization -- BL Lacertae objects: BL\,Lacertae -- BL Lacertae objects: PKS\,1424+240}

\titlerunning{Short Timescale Behavior of two BL Lacertae Type Objects}
\authorrunning{S. Covino et al.}

\maketitle

\section{Introduction}

Blazars, the subclass of active galactic nuclei (AGN) showing jets almost aligned with the observer's line of sight \citep{BR78,UP95}, offer an invaluable laboratory of physics. Their spectral energy distribution (SED) shows a characteristic double-hump shape and is usually well modeled as due to synchrotron and inverse Compton radiation \citep{Gh98}. Relativistic Doppler boosting of the observed emission is likely involved in the large-amplitude variability observed essentially at all frequencies \citep[e.g.][]{Gh93}. 

Although the interpretative scenario seems to be well established, many open problems are still present. The availability of continuously improving multi-wavelength (MW) data has revealed the need for more sophisticated approaches, with models assuming that the observed emission originates in multiple zones with typically independent physical parameters \citep[e.g.][]{Ale12}. The possibility of inhomogeneity in the emitting region, mimicked by multi-zone models, is definitely plausible. However, this immediately introduces a strong degeneracy in the already large parameter space, in turn requiring additional information to disentangle the various possible components in the observed emission.

The dominance of non-thermal emission processes (e.g. synchrotron radiation, etc.) in the blazar emission suggests that a wealth of information might come from polarimetric studies \citep[][to mention some of the most recent papers]{Lar13,Sor13,Sas14,Sor14,Zha14,Ito15}. In the optical, the detection of polarized emission was considered the smoking-gun signature for synchrotron emission from a non-thermal distribution of electrons\citep{AS80}. In general, the addition of polarimetric data to the modeling of blazar photometric/spectral information has widely shown its potential to derive information about, e.g., the magnetic field state \citep[e.g.][]{Lyu05,Mar14}, or to drive the modeling of different SED components \citep{Bar14}.

A relatively less explored regime is that of short timescale polarimetry \citep{Tom01a,Tom01b,And05,Sas08,Cha12,Ito13}. Short timescale photometry, on the contrary, is indeed a common practice in the field and has revealed to be a powerful diagnostic technique \citep{Mon06,Ran10,Dan13,Zha13,San14}, also in the very-high energy regime \citep[e.g.][]{Aha07,Alb07,Abd10,Fos13}.

In this paper we present and discuss well-sampled observations of two blazars: \object{BL\,Lacertae} (hereinafter \object{BL\,Lac}) and \object{PKS\,1424+240}. The observations were carried out  with the optical polarimeter PAOLO\footnote{\url{http://www.tng.iac.es/instruments/lrs/paolo.html}} equipping the 3.6\,m INAF / Telescopio Nazionale Galileo (TNG) at the Canary Island of La Palma. The relatively large collective area of the TNG enabled us to explore time scales as short as several tens of seconds in both photometry and polarimetry.

The paper is organized as follows: observations are described in Sect.\,\ref{sec:data}. In Sect.\,\ref{sec:resdis} results of the analyses and a general discussion are presented, and conclusions are drawn in Sect.\,\ref{sec:conc}.

\section{Observations}
\label{sec:data}

PAOLO is an optical polarimeter integrated in the Naysmith focus instrument DOLORES\footnote{\url{http://www.tng.iac.es/instruments/lrs/}} at the TNG. The observations presented here were part of the commissioning and scientific activities of the instrument. 

\object{BL\,Lac} is the prototype of the class of BL\,Lac objects, and is located at a redshift $z=0.069$ \citep{MH97}. The host is a fairly bright and massive elliptical galaxy \citep{Sca00,Hyv07}. Due to its relative proximity it is one of the most widely studied objects of the class. \object{PKS\,1424+240} is also a BL\,Lac object and its redshift is still uncertain. \citet{Fur13} report a lower limit at $z \gtrsim 0.6$, which can make it one of most luminous objects in its class. Its host galaxy was possibly detected by \citet{MR10} at typically a few percent of the nuclear emission, although \citet{Sca00} reported much fainter limits.

\begin{figure}
\centering
\resizebox{\hsize}{!}{\includegraphics{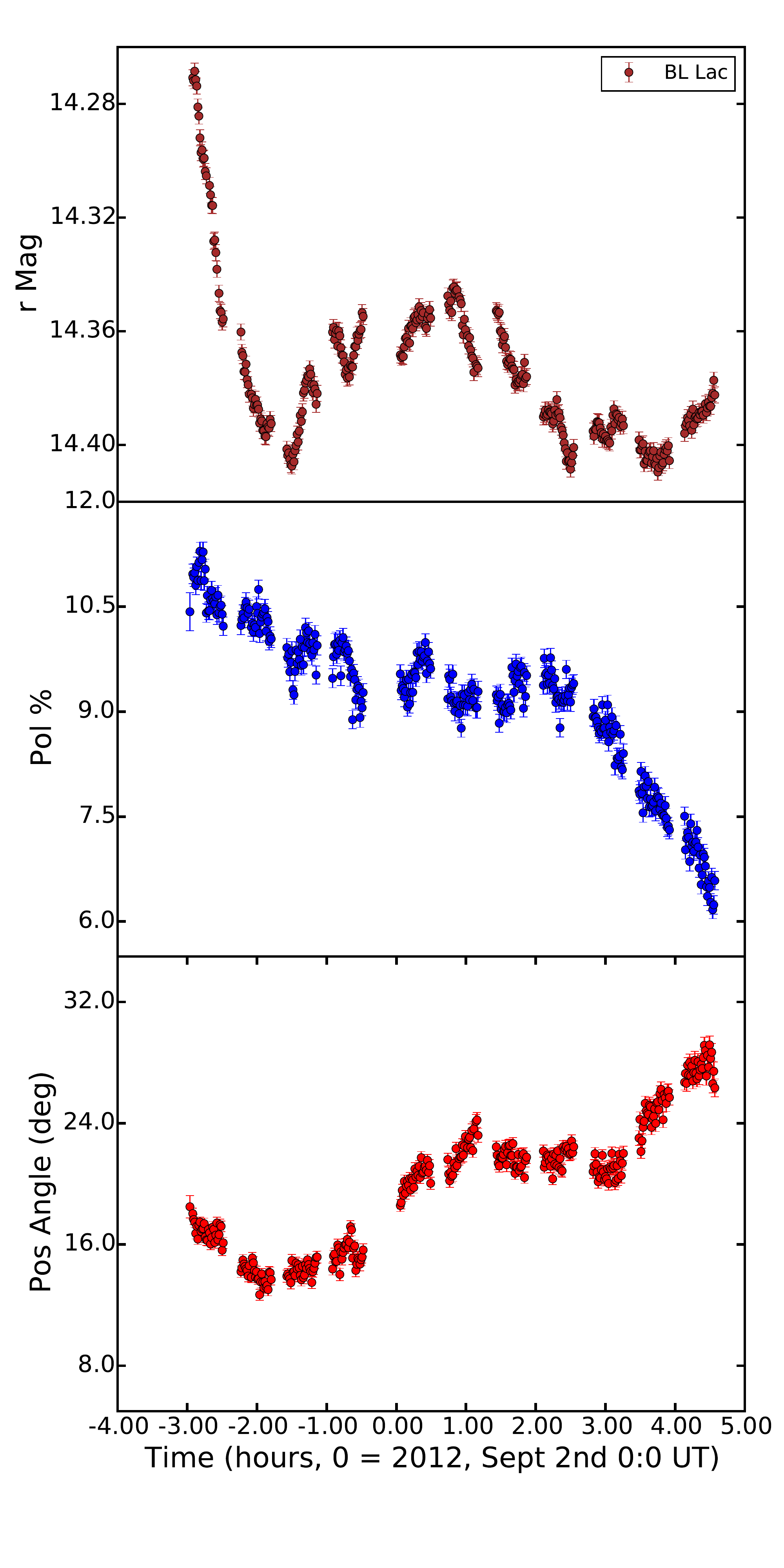}} 
\caption{PAOLO observations of BL\,Lac. In the top panel we show the magnitude of the source (AB mag) not corrected for Galactic reddening and for the host galaxy brightness. In the middle panel we show the polarization degree and in the bottom panel the position angle. }
\label{fig:bllac}
\end{figure}

\object{BL\,Lac} was observed for about 8 hours during the night of 2012 September 1 -- 2. The observations consisted of short integrations of about 20-40\,s each with the $r$ filter, interrupted every $\sim 45$\,min to observe polarized and unpolarized polarimetric standard stars (\object{BD+28d4211}, \object{W2149+021}, \object{HD\,204827}) for a total of more than 300 data points. The data reduction is carried out following standard procedures and aperture photometry is performed using custom tools\footnote{\url{https://pypi.python.org/pypi/SRPAstro.FITS/}}. Photometric calibration was secured by comparison with isolated unsaturated stars in the field with magnitudes derived by the APASS catalogue\footnote{\url{http://www.aavso.org/apass}}. Photometric and polarimetric light curves are shown in Fig.\,\ref{fig:bllac}.

\begin{figure}
\centering
\resizebox{\hsize}{!}{\includegraphics{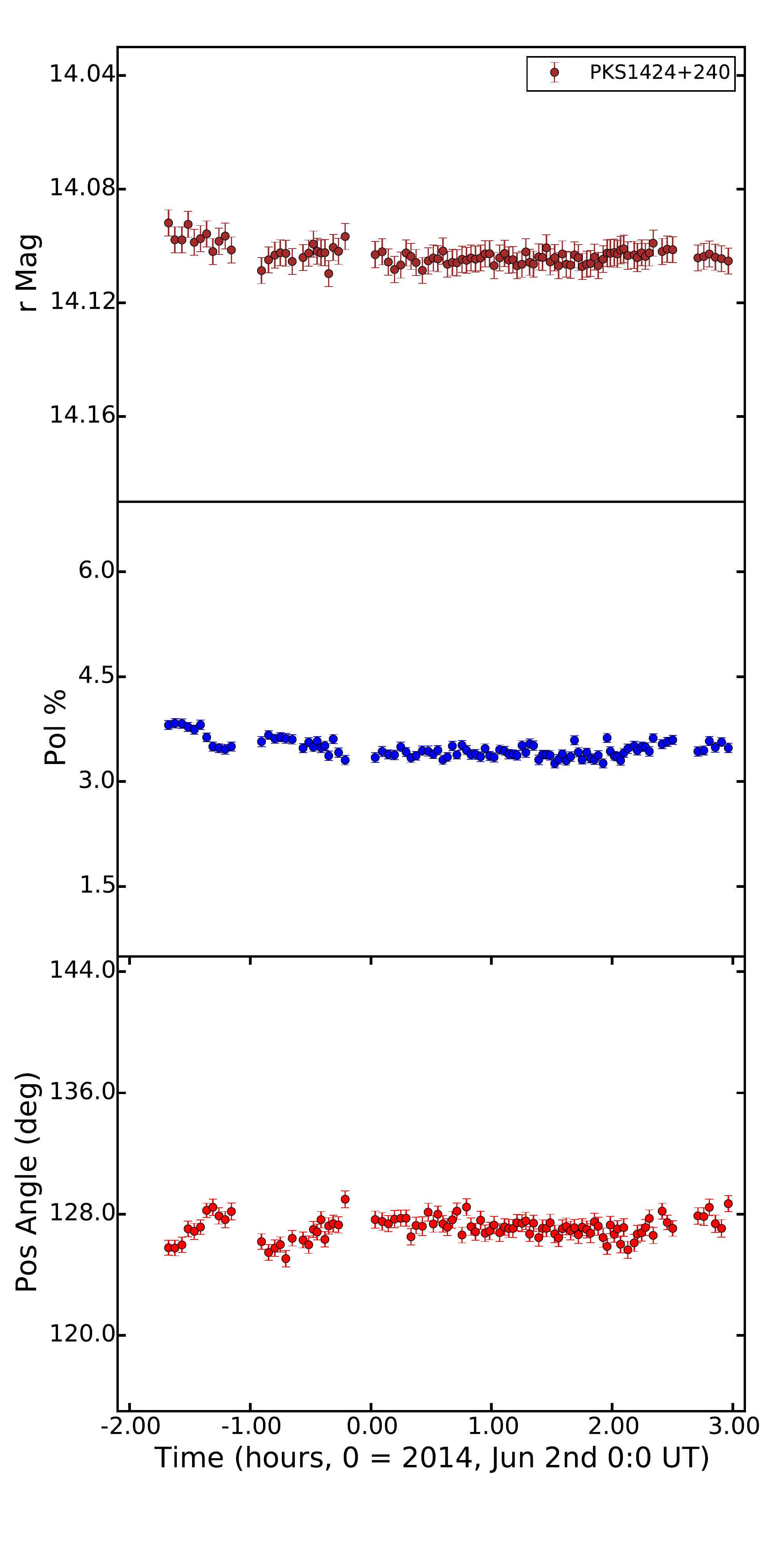}} 
\caption{PAOLO observations of PKS\,1424+240. The top panel shows the magnitude of the source (AB mag) versus time not corrected for Galactic reddening and for the host galaxy brightness. The middle panel shows the polarization degree and the bottom panel the position angle versus time.}
\label{fig:pks}
\end{figure}

\object{PKS\,1424+240} was observed for about 5 hours during the night of 2014 June 1 -- 2. The observations consisted of short integrations of 1-2\,min each with the $r$ filter interrupted at the beginning and at the end of the sequence to observe an unpolarized polarimetric standard star (GD\,319) for a total of more than 100 data points. Reduction and calibration were carried out as for \object{BL\,Lac.} Photometric and polarimetric light curves are shown in Fig.\,\ref{fig:pks}.

The removal of the few percent instrumental polarization typical of Nasmyth focus instruments \citep{Tin07,Wit11,Cov14} can be carried out rather efficiently and with PAOLO we can estimate \citep{Cov14} a residual r.m.s. of the order of $\sim 0.2$\% or better. If the observations cover a limited range in hour angles the correction is generally more accurate. This is a systematic uncertainty superposed onto our observations and it is already included in the reported errors for our data. The results reported here supersede the preliminary ones shown in \citet{Cov14}.

Where required, $\chi^2$ minimization is performed by using the downhill (Nelder-Mead) simplex algorithm as coded in the {\tt python}\footnote{\url{http://www.python.org}} {\tt scipy.optimize}\footnote{\url{http://www.scipy.org/SciPyPackages/Optimize}} library, v.\,0.14.0. The error search is carried out following \citet{Cas76}. Throughout this paper the reported uncertainties are at $1\sigma$.

Distances are computed assuming a $\Lambda$CDM-universe with $\Omega_\Lambda = 0.73, \Omega_{\rm m} = 0.27,$ and H$_0 = 71$\,km\,s$^{-1}$\,Mpc$^{-1}$  \citep{Kom11}. Magnitudes are in the AB system. Flux densities are computed following \citet{Fuk96}. The raw and reduced data discussed here are available from the authors upon request.

\section{Results and discussion}
\label{sec:resdis}

\object{BL\,Lac} and \object{PKS\,1424+240} are sources belonging to the same class and, during our observations, also showed a comparable brightness. This is already a remarkable finding since the latter is more than one order of magnitude farther away than the former. \object{PKS\,1424+240} is therefore intrinsically about 100 times more luminous in the optical than \object{BL\,Lac} in the considered period. The host galaxy of \object{BL\,Lac} was measured at $R \sim 15.5$ \citep{Sca00}, roughly 30\% of the source luminosity during our observations. The source showed intense short-term variability, as expected for a blazar which has previously been found to be strongly variable at any time-scale \citep{Rai13}. On the contrary, \object{PKS\,1424+240} was remarkably stable during the observations with slow (hours) variations at most at a few percent level. This behavior is rather unexpected although this source presented a less intense variability (at least compared to \object{BL\,Lac}) during long-term monitoring campaigns \citep[e.g.][]{Arc14,Ale14} and in particular close to our observation epoch\footnote{\url{http://users.utu.fi/kani/1m/PG\_1424+240.html}}.

\subsection{Analysis of flux variability}
\label{sec:var}

\begin{figure}
\centering
\resizebox{\hsize}{!}{\includegraphics{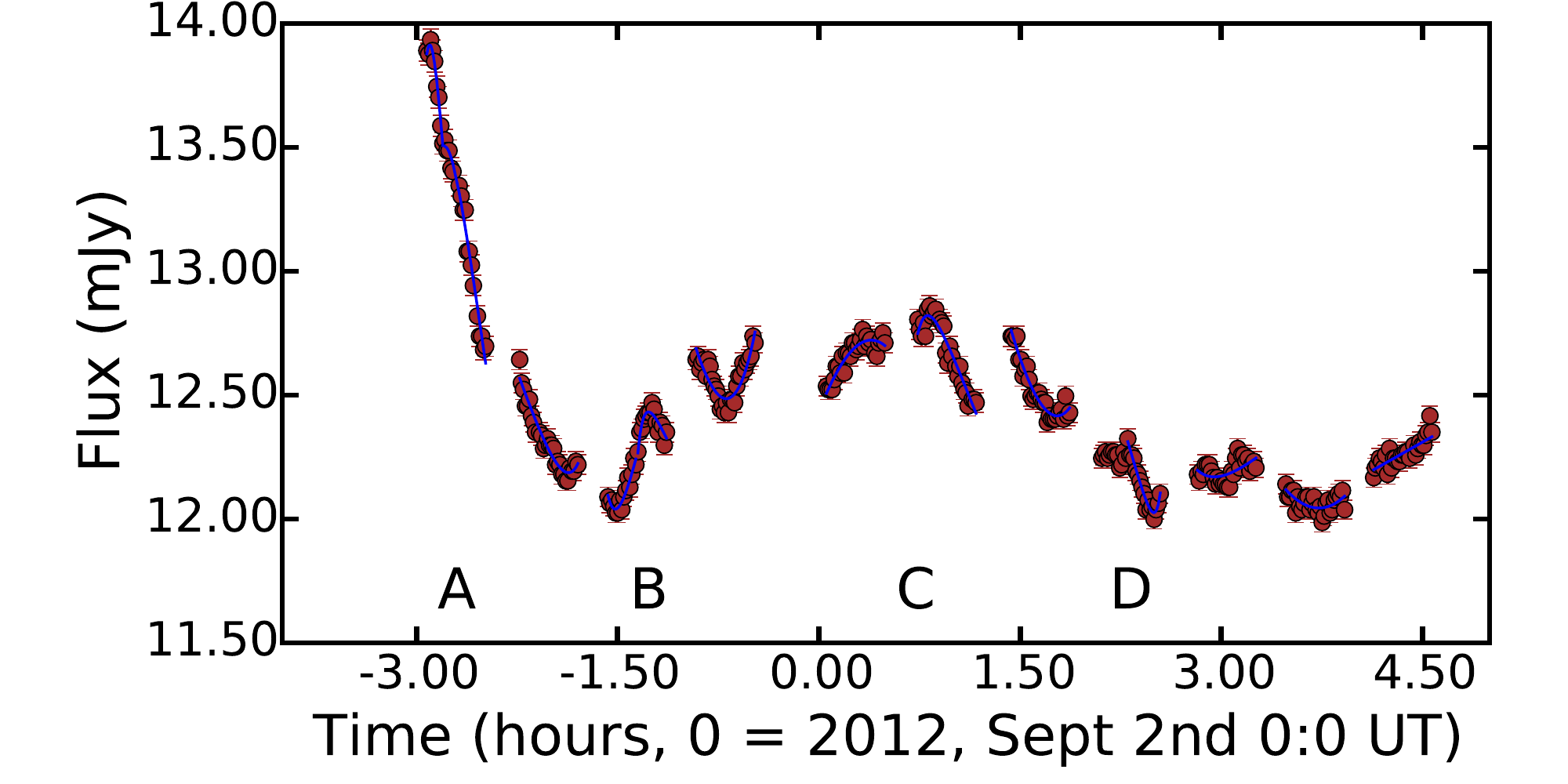}} 
\caption{BL\,Lac light-curve after subtraction of the host galaxy contribution and correction for Galactic extinction. A few episodes of rapid variability are labelled (see Table\,\ref{tab:bllacrapvar}) and fits based on Eq.\,(\ref{exppeak}) are also shown (blue solid line).}
\label{fig:bllaclc}
\end{figure}

The rapid variability observed in \object{BL\,Lac}, although in most cases of rather low level in absolute terms ($\sim 5-10$\%), is characterized by a fair number of well sampled rise/decay phases \citep[see also][for a similar behavior in S5\,0716+714]{Mon06}. Following \citet{Dan13}, we modeled these episodes with a sum of exponentials after having converted the light-curves to flux densities. The rationale is based on the idea that the derived time-scales, $\tau$, can give constraints on the size of the emitting regions. In addition, the time-scales of the decay phases, if the emission is due to synchrotron radiation, can allow us to derive inferences about the cooling times of the accelerated electrons and, in turn, the magnetic fields.

The adopted empirical functional form \citep{Dan13} is:
\begin{equation}
\label{exppeak}
f_{\rm i}(t)=\frac{2F_{\rm i}}{exp\left( \frac{t_{\rm i}-t}{\tau_{\rm {r, i}}}\right) +exp\left( \frac{t-t_{\rm i}}{\tau_{\rm {d, i}}}\right) },
\end{equation}
where $ F_{\rm i} $ is the flare normalization, $ \tau_{\rm {r, i}} $ and $ \tau_{\rm {d, i}} $ are, respectively, the flux rise and decay time-scales, and $t_{\rm i}$ is the time of the pulse maximum. The inverse of Eq.\,\ref{exppeak}, $1/f_{\rm i}(t)$, is used when the light-curve shows a decay followed by a rise, and $t_{\rm i}$ corresponds in this case to the pulse minimum.

The dense sampling of our light-curve allowed us to derive four events with well constrained time-scales (Table\,\ref{tab:bllacrapvar} and Fig.\,\ref{fig:bllaclc}). In all cases the time-scales for rise or decay phases are approximately in the range 2-15\,min, considering the uncertainties. 

Variability time-scales as short as a few minutes have already been singled out for BL\,Lac objects, mainly at high energies \citep[e.g.][]{Aha07,Alb07,Arl13} or X-rays \citep[e.g.][]{WW95}, where the flux variation is a large fraction of the total. In the optical the percentage amplitudes of flux variations are typically lower, possibly due to the superposition of several emission episodes with largely different time-scales \citep[e.g.][]{Cha11,Dan13,San14} originating from different emitting regions. Therefore, strictly speaking, constraints derived by the light-curve analysis hold only for a portion of the emitting region of the order of the ratio of the flux variability to the total flux.

\begin{table}
\caption{Parameters of rapid flares during our BL\,Lac monitoring. The epochs are relative to 00:00 UT on 2012 September 2. $F_i$ is the amplitude of the variability episode. $1\sigma$ errors are computed with two parameters of interest \citep{Cas76}.}
\label{tab:bllacrapvar}
\centering 
%\resizebox{\columnwidth}{!}{
\begin{tabular}{lrccl}
\hline \hline
Event  & Epoch & $F_i$ & $\tau$  &  notes \\
	   & (hours) & (mJy) & (min) & \\
\hline
A  &  $-2.7$ & $0.48_{-0.05}^{+0.11}$ & $3.3^{+1.2}_{-0.6}$ & decay \\
B  & $-1.3$ & $0.27_{-0.15}^{+2.80}$ & $2.5^{+17.1}_{-1.3}$ & rise \\
C  & $0.7$ & $0.13_{-0.04}^{+1.87}$ & $3.6^{+6.4}_{-1.9}$ & rise \\
D  & $2.3$ & $0.06^{+1.94}_{-0.02}$ & $2.4^{+9.1}_{-1.5}$ & decay \\
\hline
\end{tabular}
%}
\end{table}

The size of the emitting region can be constrained as:
\begin{equation}
R \lesssim \frac{\delta c \tau}{1+z},
\end{equation}
where $z$ is the source redshift, $c$ the speed of light, and $\delta$ is the relativistic Doppler factor of the emitting region. Assuming a reference time scale of $\sim 5$\,min we get $R \lesssim 3\times10^{-5} \times \frac{\delta}{10}\,{\rm pc} \sim 10^{14} \times \frac{\delta}{10}$\,cm. The rapid variability identified here amounts to only a few percent of the total emitted flux from \object{BL\,Lac}.

Under the hypothesis that the (variable) emission is due to synchrotron and Compton processes, the cooling time-scale can limit the time-scale of a decay phase as:
\begin{equation}
\label{eq:cool}
\tau_{\rm d} \gtrsim  t_{\rm cool} = \frac{3m_{e}c(1+z)}{4\sigma_{\rm T} \delta u^{'}_{0}\gamma_{e}}\,{\rm s},
\end{equation}
where $m_{\rm e}$ is the electron mass, $\sigma_{\rm T}$ the Thomson cross-section, $u^{'}_{0}=u^{'}_{B}+u^{'}_{\rm rad}=(1+q)B^{'2}/8\pi$ the co-moving energy density of the magnetic field (determining the synchrotron cooling rate) plus the radiation field (determining the inverse-Compton cooling rate), $q= u^{'}_{\rm rad}/u^{'}_{B}$ the Compton dominance parameter, typically of order of unity for BL\,Lacs \citep{Tav10}, and $\gamma_{e}$ is the characteristic random Lorentz factor of electrons producing the emission.

The peak frequency of the synchrotron emission is at
\begin{equation}
\label{eq:syn}
\nu_{\rm syn}=\frac{0.274 \delta e \gamma_{e}^{2}B^{'}}{(1+z)m_{e}c}\,{\rm Hz},
\end{equation}
where $e$ is the electron charge.

Finally, substituting $\gamma_e$ in Eq(s).\,\ref{eq:cool} and \ref{eq:syn}, the co-moving magnetic field can be constrained as:
\begin{eqnarray}
B^{'} \gtrsim [\pi m_e c (1+z) e / \sigma_{\rm T}^{2}]^{1/3}  \nu_{\rm syn}^{-1/3} t_{\rm cool}^{-2/3} \delta^{-1/3} \sim \\ \nonumber
\sim 4 \times 10^7 (1+z)^{1/3} \nu_{\rm syn}^{-1/3} t_{\rm cool}^{-2/3} \delta^{-1/3}\,{\rm G}.
\end{eqnarray}

\object{BL\,Lac} is an intensively monitored object. \citet{Rai13} reported on a comprehensive study of its long-term behavior, including the epoch of our observations. From that data set the position of the synchrotron peak frequency can be inferred to be close to $\nu_{\rm syn} \sim 5 \times 10^{14}$\,Hz, and therefore, again assuming a reference time scale for decay of $\sim 5$\,min, and considering that the cooling time should be shorter than this,  we get $B' \gtrsim 6 \times (\frac{\delta}{10})^{-1/3}$\,G.

The SED of \object{BL\,Lac} and that of a number of sources of the same class were studied in \citet{Tav10} based on observations carried out in 2008. A single zone model allowed the authors to estimate an average magnetic field $B \sim 1.5$\,G, a Doppler factor $\delta \sim 15$, and radius of the emitting region $R \sim 7 \times 10^{-4}$\,pc. Compared to the results from our analysis, based however on observations carried out in 2012, the emitting region of \object{BL\,Lac} turns out to be, as expected, a small fraction of that responsible for the whole emission and the magnetic field is locally higher but still close to the one zone model inference. 

A similar analysis for \object{PKS\,1424+240} is not possible due to the very low level of variability shown during our observations. A fit with a constant is indeed perfectly acceptable although during the first $\sim 30$\,min of observations the source was slightly brighter by $\sim 0.01-0.02$\,mag.

The length of our monitoring does not allow us to derive general conclusions, although the difference in the observed flux variability between the two objects is remarkable. \object{PKS\,1424+240} is actually at a higher redshift compared to \object{BL\,Lac} ($z \sim 0.6$ vs. $z = 0.069$). Time dilation will lead to a reduction of any intrinsic variability for the former source by a factor of about 1.5 with respect to the latter. In addition, based on the SEDs shown in \citet{Tav10} and \cite{Ale14}, the optical band is at a higher frequency than the synchrotron peak for \object{BL\,Lac}, and at a lower frequency (or close to) for \object{PKS\,1424+240}. As widely discussed in \citet{Kir98}, under the assumption that magnetic fields in the emitting region are constant, flux and spectral variability depend on the observed frequency. If electrons with a given energy, corresponding to photons at a given frequency, cool more slowly than they are accelerated, variability is smoothed out, as it might be the case for \object{PKS\,1424+240}. Variability is expected to be particularly important close to frequencies emitted by the highest-energy electrons, where both radiative cooling and acceleration have similar timescales. Different short-term variability behaviors for sources with the synchrotron peak at lower or higher frequencies than the observed band were indeed already singled out \citep{HW96,HW98,Rom02,Hov14}.

In the literature it is also customary to look for the total flux doubling/halving times \citep[e.g.][]{Sba11,Imp11,Fos13}. The small amplitude of the variability we observed does not allow us to derive strong constraints, since this would always require large extrapolation. However, the shortest time-scales we could detect are of the order of less than four hours for \object{BL\,Lac}, consistent with the values found in other blazars, which is consistent with the idea that the the whole emitting region is much larger than the regions responsible for the rapid variability.

A variability analysis can be carried out for the polarimetric light curves too. The results show variability timescales at the same level as the total flux curves, although with larger uncertainties. Rapid time variability on minute to hours time scales for the polarized flux was singled out in other blazars, as for instance AO\,0235+164 \citep{Hag08}, S5\,0716+714 \citep{Sas08}, CGRaBS\,J0211+1051 \citep{Cha12} or CTA\,102 \citep{Ito13}. Intranight variability for a set of radio-quiet and radio-loud AGN was studied by \citet{Vil09}.

\subsection{Polarimetry}

Blazar emission is known to be characterized by some degree of polarization that is often variable, both in intensity and direction, on various time-scales \citep[see][for a recent review about optical observation of BL\,Lacs]{Fal14}. Occasionally, some degree of correlation or anticorrelation between the total and polarized flux is observed \citep[e.g.][]{Hag08, Rai12,Sor13,Gau14}, while often no clear relation is singled out. The complexity of the observed behaviors likely implies that, even when a single zone modeling can satisfactorily describe the broad-band SEDs, more emission components are actually active. It was proposed \citep[e.g.][]{Bar10,Sak13} that a globally weakly polarized fraction of the optical flux is generated in a relatively stable jet component, while most of the shorter term variability, both in total and polarized flux, originates from the development and propagation of shocks in the jet. 

\object{BL\,Lac} and \object{PKS\,1424+240} show rather different behaviors in the linear polarimetry as well. The degree of polarization of \object{BL\,Lac} starts at about 11\% and decreases slowly for a few hours to about 9\%; then, for the remaining three hours of our monitoring, it decreases more quickly to about 6\%. The position angle increases rather quickly after the first hour, from about $14^\circ$ to $23^\circ$; then it remains stable for a couple of hours and then increases again to about $30^\circ$. Superposed on these general trends there is considerable short-term variability above the observational errors. \object{PKS\,1424+240}, on the contrary, shows a fairly constant polarization degree at about 4\% and a position angle close to $127^\circ$, with some variability only at the beginning of our monitoring. These behaviors are in general agreement with the results reported by \citet{And05} studying intra-night polarization variability for a set of BL\,Lac objects.

The \object{PKS\,1424+240} jet was likely in a low activity state, although it was not in its historical minimum (see Sect.\,\ref{sec:resdis}). This is also confirmed by the publicly available information and data at other wavelengths, such as high-energy gamma rays provided by the {\it Fermi}/LAT Collaboration\footnote{{\tt http://fermisky.blogspot.it/2014\_06\_01\_archive.html}}, and soft X-rays available from the {\it Swift}/XRT monitoring program\footnote{{\tt http://www.swift.psu.edu/monitoring/source.php?source=PKS1424+240}}. \citet{Ale14} reported a higher polarization degree, $7-9$\%, in 2011, when the source was brighter than during our monitoring. Lower polarization degrees, $4.4-4.9$\%, were reported by \citet{Mea90} in 1988, when the source was instead fainter. The position angle was about $113-119^\circ$, similar to that observed during our monitoring. The latter is also consistent with the direction of the jet as measured by VLBA radio observations at 2\,cm (Lister et al. 2013). The kinematics of the most robust radio component showed a position angle of $141^{\circ}$ with a velocity vector direction of $108^{\circ}$, i.e. with a very small offset \citep[$33^{\circ}\pm20^{\circ}$,][]{Lis13}. VLBA observations in the framework of the MOJAVE Project\footnote{{\tt http://www.physics.purdue.edu/astro/MOJAVE/sourcepages/1424+240.shtml}}  \citep{Lis09} showed a decreasing trend in the polarization degree from 5\% in 2011 to 2.8\% in 2013, with a roughly stable position angle ($126^{\circ}-154^{\circ}$), consistent with our results in the optical. In general, looking at historical data, the polarization degree of  \object{PKS\,1424+240} seems to be almost constant ($\sim 4$\%) below a given optical flux (likely $\lesssim 9.0$\,mJy, based on the refereed studies). The optical position angle seems to be quite stable and aligned with the kinematic direction of the radio jet and with the radio polarization position angle. This behavior might suggest some kind of ``magnetic switch'' (i.e. a threshold effect) in the jet activity \citep[e.g.][]{PuCo90, Mei97, Mei99}.

Neglecting the short-term variability of  \object{BL\,Lac}, the total rotation of the position angle, taking the minimum at approximately $\sim -2$\,hours (see Fig.\,\ref{fig:bllac}) and the value at the end of our monitoring, amounts to about $15^\circ$, i.e. $2 - 2.5^\circ$/hour ($45-60^\circ$/day). Rapid position angle rotations of this magnitude are not unusual for blazars in general, and for \object{BL\,Lac} specifically \citep[e.g.][]{All81,Sil93,Mar08}. The observation of relatively stable and long-lasting rotational trends (days to months) suggested that the polarized emission could be generated in a jet with helical magnetic fields or crossed by transverse shock waves, or in a rather stable jet with an additional linearly rotating component \citep{Rai13}.

\begin{figure*}
\centering
\resizebox{\hsize}{!}{\includegraphics{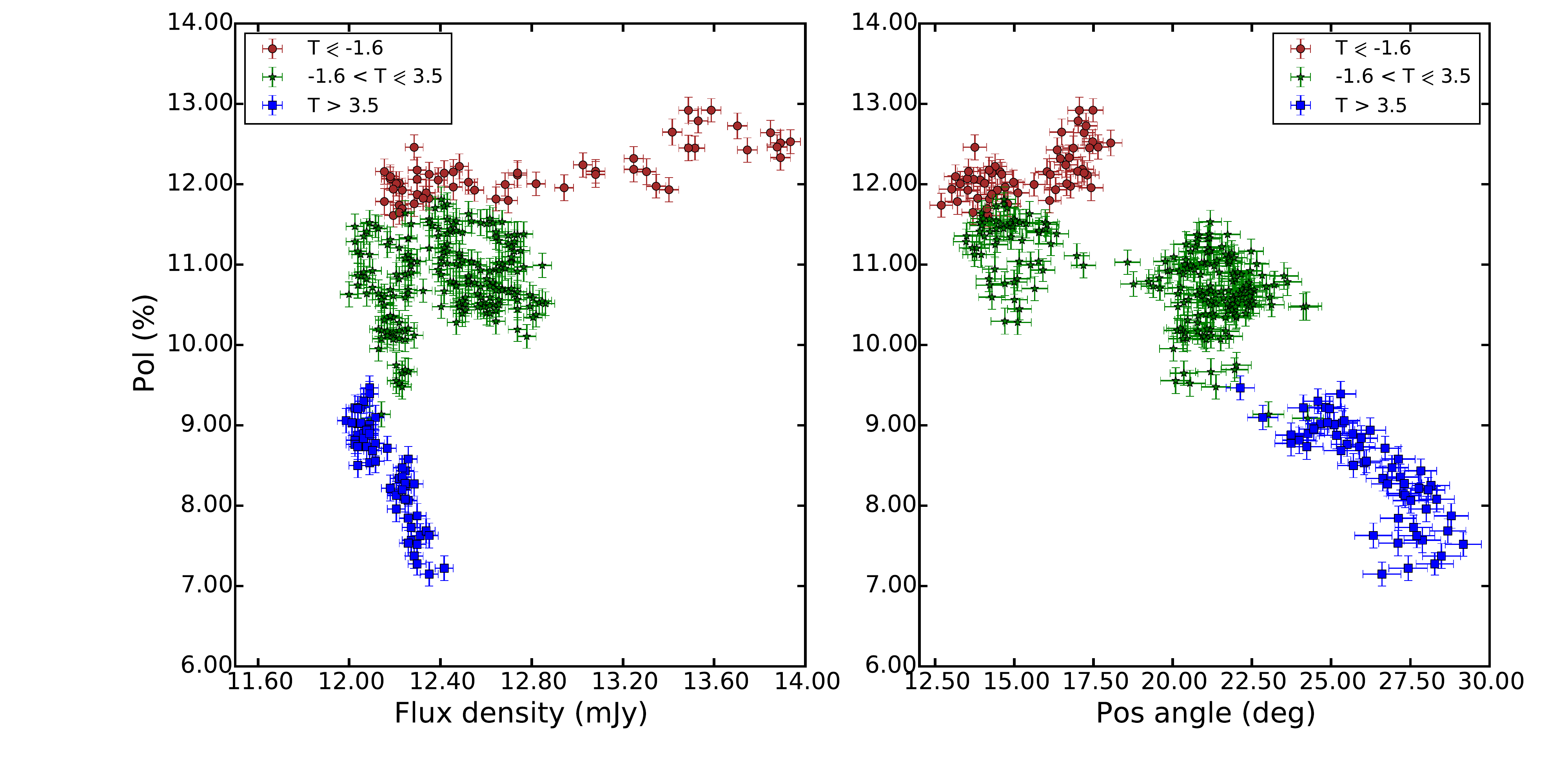}} \\
\resizebox{\hsize}{!}{\includegraphics{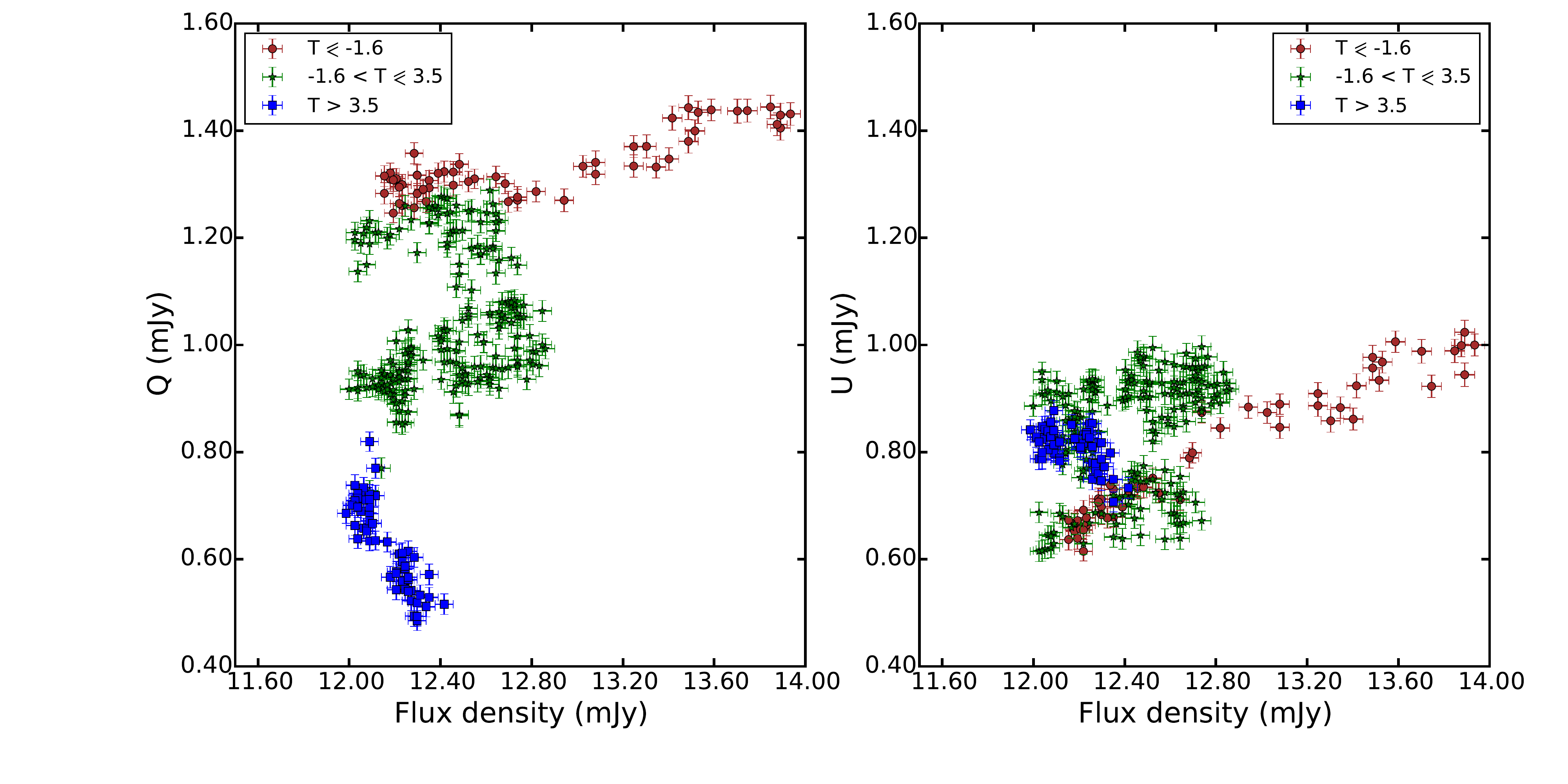}}
\caption{({\it upper left}) BL\,Lac (host galaxy subtracted) flux density vs. linear polarization \citep[host galaxy corrected, assuming unpolarized emission, e.g.][]{Cov03}. At least three different regimes are singled out: at early time the flux changes rapidly with a slowly varying and rather high polarization (brown, circles), then an intermediate phase with chaotic flux and polarization variations (green, stars), and finally a sharp decrease in polarization with almost constant flux (blue, squares). Times in the legend are in hours (see Fig.\,\ref{fig:bllac}). ({\it upper right}) BL\,Lac position angle vs. linear polarization. Same symbols of the upper left panel. The position angle tends to increase when the linear polarization decreases. The trend becomes very clear at the end of our observation. ({\it bottom}) Flux density vs. Stokes parameters $Q$ and $U$. Periods with approximately linear dependence between polarization and total flux are also singled out. Same symbols as in the upper left panel.}
\label{fig:bllacpol}
\end{figure*}

Our well-sampled monitoring observations allow us to disentangle different behaviors even during the relatively short-duration coverage of \object{BL\,Lac} (Fig.\,\ref{fig:bllacpol}, upper left plot). At the beginning of our monitoring period, we see a rapid flux decrease with polarization slowly decreasing. After that, the source enters a phase characterized by rapid small-scale variability both in the total and polarized flux. Finally, the flux begins to increase regularly by a small amount and the polarization decreases abruptly down to the lowest observed level. The relation between polarization and position angle (Fig.\,\ref{fig:bllacpol}, upper right plot) shows the already mentioned rotation of the position angle with the decrease of the linear polarization. However, again superposed on this general trend there is considerable variability \citep[see also][for a similar analysis]{Hag08}.

In \citet{Rai13} the long-term (years) flux light curve was modeled assuming the flux variation to be (mainly) due to Doppler factor variations with a nearly constant Lorentz factor, i.e. due to small line of sight angle variations. We applied the same technique for our rapid monitoring. Knowing the viewing angle required to model the flux variations, it is then possible to predict the expected polarization in different scenarios. In the case of helical magnetic fields, following \citet{Lyu05}, we can derive a polarized flux fraction at 9-10\%, roughly in agreement with our observations. However, a detailed agreement, explaining the short-time variability for both the total and polarized flux, is not possible. Alternatively, we may consider transverse shock wave models \citep{Hug85}, with which again rough agreement for the polarization degree is reached, but no detailed agreement is possible. 
A geometric model for the flux variation is therefore unable to simultaneously interpret the total flux and polarization behavior at the time resolution discussed here.

As already introduced in Sect.\,\ref{sec:var}, a possible interpretation of both total and polarized flux curves can be derived if it is assumed that the observed emission is due to a constant (within the time-scale of our monitoring) component with some degree of polarization and one \citep[or many,][]{Bri85} rapidly varying emission component(s) with different polarization degree and position angle \citep[see also][]{Sas08,Sak13}. The idea is rather simple; using the first three Stokes parameters the observed polarization can be described as:
\begin{equation}
S = \left \{
\begin{aligned}
I_{\rm obs} & = & I_{\rm const} & + & I_{\rm var} \\
Q_{\rm obs} I_{\rm obs} & = & Q_{\rm const} I_{\rm const} & + & Q_{\rm var} I_{\rm var} \\
U_{\rm obs} I_{\rm obs} & = & U_{\rm const} I_{\rm const} & + & U_{\rm var} I_{\rm var}
\end{aligned}
\right .
\label{eq:stokes}
\end{equation}
where the suffixes "obs", "const" and "var" refer to the observed (total), constant and variable quantities. The redundancy in Eq.\,\ref{eq:stokes} can be reduced following various possible assumptions, often depending on the availability of multi-wavelength datasets or long-term monitoring \citep[see, e.g.][]{Hol84,Qui93,Bri96,Bar10}. \citet{Hag02} assumed, based on their long-term polarimetric monitoring, that the stable component in \object{BL\,Lac} could be characterized by $P \sim 9.2$\% and $\theta \sim 24^\circ$. 
%With these parameters, and assuming indicatively $I_{\rm const} \sim 8$\,mJy, the variable component, which is responsible only for a fraction of the total flux, is highly polarized, up to about 50\%, in qualitative agreement with the results discussed in \citet{Hag02} but on a much shorter time-scale. 
As discussed in \citet{Hag99} and \citet{Hag08}, if a linear relation between polarized and total flux is singled out, this can allow one to estimate the polarization degree and position angle of the variable component. A linear relation between the Stokes parameters and the total flux implies that polarization degree and position angle are essentially constant \citep{Hag99} and their values can be derived as the slopes of the linear relations. At the beginning of our monitoring we can identify a sufficiently long and well defined linear relation between the Stokes parameters $Q$ and $U$ and the total flux (see Fig\,\ref{fig:bllacpol}, bottom panel). As already mentioned, we find considerable variability superposed on the linear trend. Neglecting the shorter term variability, we can roughly estimate $P_{\rm var} \sim 22$\% and $\theta \sim 34^\circ$. The constant component turns out to be remarkably consistent with the one identified by \citet{Hag02} at a flux level $\sim 9.5$\,mJy.

At any rate, the increasing complexity singled out by long- and short-term monitoring requires new theoretical frameworks for a proper interpretation.  \citet{Mar14}, for instance, proposed a scenario in which a turbulent plasma is flowing at relativistic speeds crossing a standing conical shock. In this model, total and polarized flux variations are due to a continuous noise process rather than by specific events such as explosive energy injection at the basis of the jet. The superposition of ordered and turbulent magnetic field components can easily explain random fluctuations superposed on a more stable pattern, without requiring a direct correlation between total and polarized flux. As discussed in \citet{Mar14}, simulations based on this scenario can also give continuous and relatively smooth position angle changes as observed during our monitoring of \object{BL\,Lac}.

\begin{table}
\caption{Parameters of interest for the model based on the scenario extensively discussed in \citet{Zha14} and \citet{Zha15}. Angles are in the co-moving frame.}
\label{tab:bllacsim}
\centering 
\begin{tabular}{lc}
\hline \hline
Parameter  & Value \\
\hline
Bulk Lorentz factor   & 15 \\
%Length of the emission region ($Z$) & $1.13\times10^{16}$\,cm \\
%Radius of the emission region ($R$) & $5.60\times10^{15}$\,cm \\
Length of the disturbance ($L$) & $3.8\times10^{14}$\,cm \\
Radius of the disturbance ($A$) & $4.0\times10^{15}$\,cm \\
Orientation of the line of sight & $90^\circ$ \\
%Electron acceleration time-scale ($Z/c$) & $9.00\times10^{-2}$\,s \\
%Electron escape time-scale ($Z/c$) & $1.50\times10^{-2}$\,s \\
%Electron background temperature &	$1.00\times10^{2}$\,m$_{\rm e}$c$^2$k$_{\rm B}^{-1}$ \\
Helical magnetic field strength & 2.5\,G \\
Helical pitch angle & $47^\circ$ \\
Electron density & $4.5\times10^{2}$\,cm$^{-3}$ \\
%Electron acceleration time-scale during disturbance $(Z/c)$ &	$4.50\times10^{-2}$\,s \\
\hline
\end{tabular}
\end{table}

\citet{Zha14} presented a detailed analysis of a shock-in-jet model assuming a helical magnetic field throughout the jet. They considered several different mechanisms for which a relativistic shock propagating through a jet may produce a synchrotron (and high-energy) flare. They find that, together with a correlation between synchrotron and synchrotron self-Compton flaring, substantial variability in the polarization degree and position angle, including large position angle rotations, are possible. This scenario assumes a cylindrical geometry for the emitting region moving along the jet, which is pervaded by a helical magnetic field and a turbulent component. On its trajectory, it encounters a flat stationary disturbance, which could be a shock. This shock region does not occupy the entire layer of the emitting region, but only a part of it. In the comoving frame of the emitting region, this shock will travel through the emitting region, and temporarily enhance the particle acceleration, resulting in a small flare. After the shock moves out, the particle distribution will revert to its initial condition due to cooling and escape. 

\begin{figure}
\centering
\resizebox{\hsize}{!}{\includegraphics{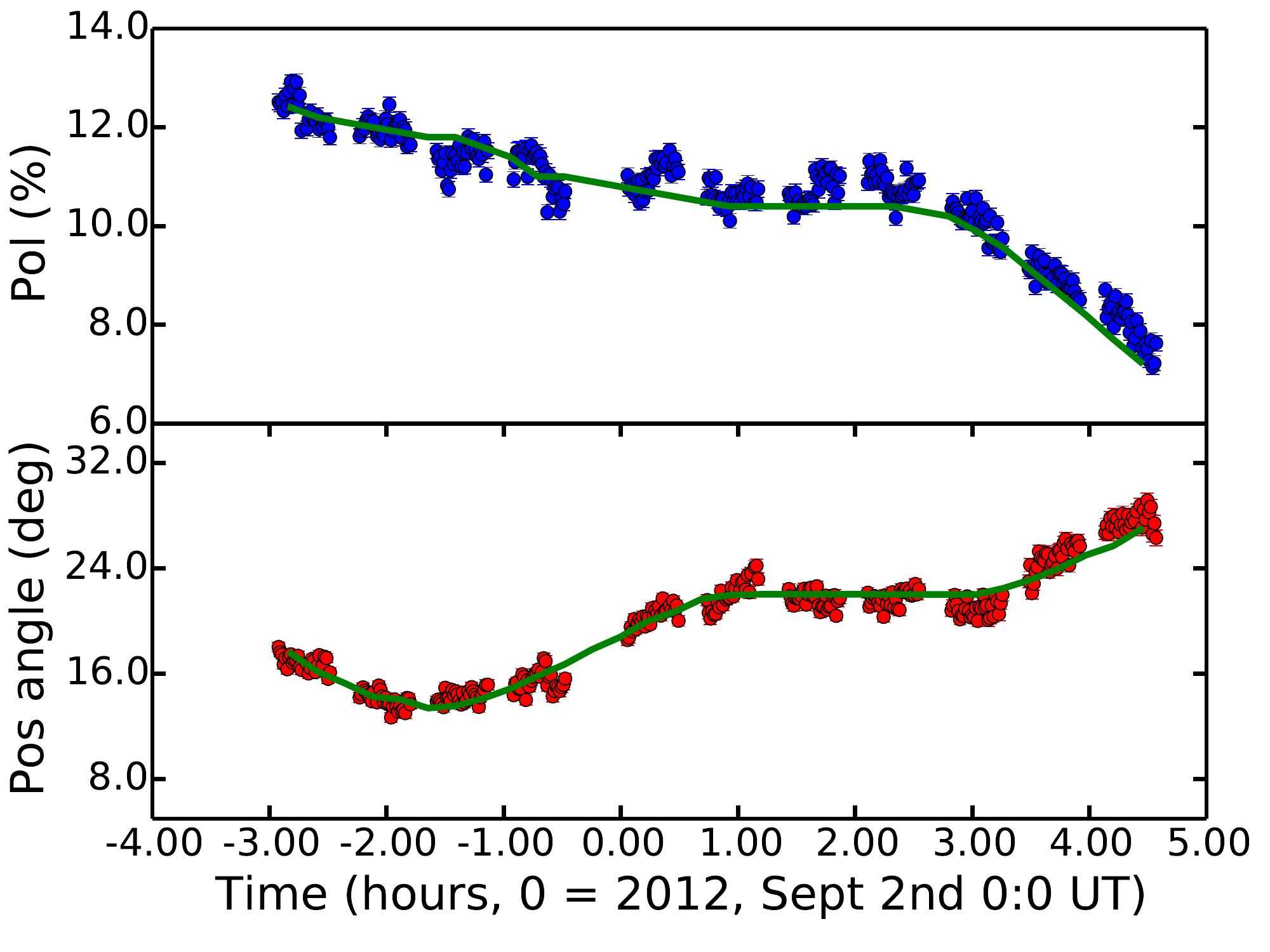}} 
\caption{Fit of the \citet{Zha14,Zha15} model scenario (green solid line) to the \object{BL\,Lac} linear polarization \citep[host galaxy corrected, assuming unpolarized emission, e.g.][]{Cov03} and position angle data. The general behavior is fairly well described by the model, although for the polarization the addition of weakly constrained turbulence in the magnetic field is required.}
\label{fig:bllacmod}
\end{figure}

The 3DPol (3D Multi-Zone Synchrotron Polarization) code presented in \citet{Zha14} and MCFP (Monte Carlo/Fokker-Planck) code presented in \citet{Che12} realizes the above model. As elaborated in \citet{Zha15}, since the shock is relatively weak and localized, the enhanced acceleration will lead to a small time-symmetric perturbation in the polarization signatures. Some of the key parameters for the model are reported in Table\,\ref{tab:bllacsim}, and the fits to the polarization degree and position angle light curves are shown in Fig. \ref{fig:bllacmod}. Near the end of the observation, the polarization degree experienced a sudden drop, while the position angle continued to evolve in a time-symmetric pattern. Therefore an increase in the turbulent contribution is necessary although, due to the lack of a multi-wavelength SED, it cannot be well constrained. The total flux, given the very low variability amplitude observed during our observations, was set at a constant level of about 12\,mJy (see Fig.\,\ref{fig:bllaclc}). Nevertheless, rapid polarimetry clearly reveals its diagnostic power, showing need of inhomogeneity and turbulence in the emitting region.

\section{Conclusions}
\label{sec:conc}

In this work we are presenting results from rapid time-resolved observations in the $r$ band for two blazars: \object{BL\,Lac} and \object{PKS\,1424+240}. The observations were carried out at the 3.6\,m TNG and allowed us to measure linear polarimetry and photometry almost continuously for several hours for both sources. In practice, long-term monitoring observations of relatively bright blazars can only be achieved with dedicated small-size telescopes; however the richness of information obtainable with a rather large facility as the TNG allows us to study regimes which were in the past only partially explored.

\object{BL\,Lac} and \object{PKS\,1424+240} show remarkably different variability levels, with the former characterized by intense variability at a few percent level, while the latter was almost constant for the whole duration of our observations. The shortest well constrained variability time scales for \object{BL\,Lac} are as short as a few minutes, allowing us to derive constraints on the physical size and magnetic fields of the source regions responsible for the variability. 

The variability time-scales for the polarization of \object{BL\,Lac} are compatible with those derived for the total flux, while \object{PKS\,1424+240} shows an almost constant behavior also in the polarization. The position angle of \object{BL\,Lacertae} rotates quasi-monotonically during our observations, and an analysis of the total vs. polarized flux shows that different regimes are present even at the shortest time-scales. 

Different recipes to interpret the polarimetric observations are considered. In general, with the simplest geometrical models, only the average level of polarization can be correctly predicted. More complex scenarios involving some turbulence in the magnetic fields are required, and promising results are derived by a numerical analysis carried out following the framework described in \citet{Zha14,Zha15}, which requires some symmetry in the emitting region, as shown by the time-symmetric position angle profile. The time-asymmetric polarization profile, and its decrease during the second part of the event, which is accompanied by a few small flares, can be described by adding some turbulent magnetic field structure to the model.

\begin{acknowledgements}
This work has been supported by ASI grant I/004/11/0. HZ is supported by the LANL/LDRD program and by DoE/Office of Fusion Energy Science through CMSO. Simulations were conducted on LANL's Institutional Computing machines. The work of MB is supported by the Department and Technology and the National Research Foundation of South Africa through the South African Research Chair Initiative (SARChI)\footnote{Any opinion, finding, and 
conclusion or recommendation expressed in this material is that of the authors and the NRF does not accept any 
liability in this regard.}. We also thank the anonymous referee for her/his competent comments that greatly enhanced the quality of the paper.
\end{acknowledgements}

%-------------------------------------------------------------------

\end{document}